\documentclass[12pt,twocolumns]{iopart}

\usepackage{iopams}
\usepackage{epsfig}
\begin{document}

\title{The Scale of Cosmic Isotropy}
\author{C. Marinoni$^{1,2}$, J. Bel$^{1}$ and A. Buzzi$^1$}

\address{$^1$ Centre de Physique  Th\'eorique,  Aix-Marseille Universit\'e, CNRS UMR 7332, case 907, F-13288 Marseille, France\\
$^2$ Institut Universitaire de France, 103, bd. Saint-Michel, F-75005 Paris, France}

\ead{christian.marinoni@cpt.univ-mrs.fr}

\begin{abstract}
The most fundamental  premise to the standard  model of the universe 
states that  the large-scale properties of the universe are the same 
in all directions and at all comoving positions. Demonstrating this 
hypothesis has proven to be a formidable challenge.
The cross-over scale $R_{iso}$  
above which the galaxy distribution becomes statistically isotropic  is vaguely defined  and  poorly  (if not at all) quantified.  
Here we report on a formalism that allows us to provide an unambiguous operational definition 
and an estimate of $R_{iso}$. We apply the method to galaxies in the Sloan Digital 
Sky Survey (SDSS) Data Release 7, finding that $R_{iso}\sim 150h^{-1}$Mpc.  Besides providing a consistency test of the Copernican principle,
 this result is in agreement with predictions based on numerical simulations of the spatial distribution of galaxies in cold dark matter dominated cosmological models.

\end{abstract}


\maketitle

\section{Introduction}
\label{intro}

The cosmological principle (CP), the assertion that the cosmic mass 
distribution appears homogeneous and isotropic, that is  uniform,  
to a family of typical observers that move with the same average velocity of the surrounding matter (fundamental or comoving observers)  has far 
reaching consequences in cosmology \cite{wey72}. It entails that 
the geometry of space-time be highly symmetric and completely described by the 
simple  Robertson-Walker metric \cite{Robe,Walk}.
Furthermore,  it implies that space expands at a rate that is set by  
the equations of Friedmann \& Lemaitre \cite{Fried,Lemai}.

The very first surveys  of the three-dimensional distribution of optical galaxies 
showed that the  topology of the large-scale structure is  very complex and irregular  \cite{gel, GH}. 
Because of this departure from exact uniformity, the  CP is regarded as a coarse-grained model of the universe,  a statistical description 
of the mass distribution that applies only on sufficiently large scales where the finest details of the galaxy clustering
pattern become  irrelevant.

More recently,  two-dimensional observations of the Cosmic Microwave Background (CMB)
\cite{ben} have shown that the universe  is extremely isotropic about us 
(to roughly  1 part in 10$^5$),  confirming earlier claims based on the analysis of the 
spatial distribution of local ($z\sim1$) sources (e.g. \cite{con}).
What is challenging is to show that the universe is isotropic also about 
distant observers. As difficult as it may seem, it is important to attack the problem.
Indeed, while isotropy at a specific position
does not imply cosmic homogeneity (and {\it viceversa}), 
isotropy about every fundamental observer does imply overall homogeneity \cite{eh}.
Lacking direct evidence for everywhere isotropy, the case for the CP rests more on 
philosophical rather than on empirical evidence (\cite{ellis, clar});   it is enough to postulate that we are not privileged observers (the 
so called Copernican principle) to deduce that if the universe appear isotropic 
about our position, it must also appear  isotropic to observers in other galaxies.

The tremendous explanatory power of the standard model of cosmology 
cannot be advocated as an indirect demonstration of the CP, since the
Friedmann-Lemaitre-Robertson-Walker (FLRW) models
are not the only solutions of the Einstein equations which are able to fit 
cosmological observations. In particular, many authors have speculated that 
some effects of the accelerated expansion of the universe \cite{de1, de2, de3, de4, de5}, which 
remains fundamentally unexplained in terms of microscopic physics, could be mimicked 
by allowing violations of  the CP  (e.g. \cite{clamaa, inho}).  
This intriguing possibility has motivated recent attempts
of rooting the CP on a more solid basis. 
Interestingly, there are some encouraging  proposals in this direction which are 
based on the analysis of the large-scale maps
of CMB anisotropies \cite{cbr, cbr2, cbr3, cbr4}, of galaxies \cite{gal, gal2, gal3, gal4} and of supernovae \cite{sn}.

Even if we postulate the CP, the picture is not complete unless we identify the  
averaging scale that is implicit in this assumption, i.e.\ the scale on which the
 FLRW model provides an effective, coarse-grained description of the universe 
\cite{smell}. It is generically asserted that the CP holds on domains that 
are large enough to encompass the biggest gravitational structures of the universe. Yet, few studies have attempted to narrow in on the length value  above which clumpiness 
gives way to uniformity \cite{lahav}.

Past efforts were mostly based on the analysis of the two-point correlation 
properties of galaxy samples \cite{davguz, davguz2}. This approach, however, 
suffers from severe theoretical drawbacks. Since  the average number density of the sample 
is needed as input,  the method  presupposes the premise to be tested,  
i.e.\ a  constant density  distribution of  matter \cite{pietro}. Moreover,   
it  does not provide an unambiguous definition of the  
cross-over scale \cite{gaite}. As a consequence, the inferred
homogeneity length-scales depend on the  size of the analyzed sample 
and range from values as 
low as $30h^{-1}$Mpc up to 200$h^{-1}$Mpc \cite{corr, corr2, corr3, corr4, corr5, corr6}.
More recently, orthogonal techniques have been explored  which are based on 
the count-in-cells analysis of  observations confined to a spatial hyper-surface
of constant time (e.g. \cite{bagla}). 
These methods are insensitive to light cone effects,  
i.e.\  possible biases  arising from  comparing galaxy fluctuations at different 
cosmic epochs \cite{maart}, and seem to indicate a transition to homogeneity at a 
scale of 70$^{-1}$Mpc \cite{cic, cic2} (but see \cite{syl} for an opposite conclusion).
In  particular, the counting method advocated by \cite{scr} allows to estimate the 
homogeneity scale independently from the sample size.

It is widely believed that, since we cannot point telescopes from 
any other place but the solar system, it is not possible to establish 
if also distant observers see an isotropic universe.  
While this argument  is certainly true for apparent 2D quantities such as, 
for example, the CMB  temperature (but see \cite{cbr}), we show here 
that it does not apply to 3D maps of the spatial distribution of galaxies.
Specifically, we quantify the typical dimension  above which independent 
observers see an isotropic  `bath' of galaxies.  
Besides establishing an operational definition of the isotropy scale, our approach  
also provides an overall consistency test for the hypothesis
that we are not privileged observers of the universe.

\section{The Method}


We identify paths of extremal length radiating 
from a given arbitrary target galaxy to every other $n^{th}$ closest neighbour (see Figure 1). 
The amplitude of the angle $t$ between these directions and the observer line-of-sight 
({\it los}) to the target  is computed  by assuming that the local properties of a homogeneous and isotropic universe are described in terms of the infinitesimal
Robertson-Walker \cite{Robe,Walk} line element 

\begin{figure}[ht]
\centerline{\includegraphics[width=80mm,angle=0]{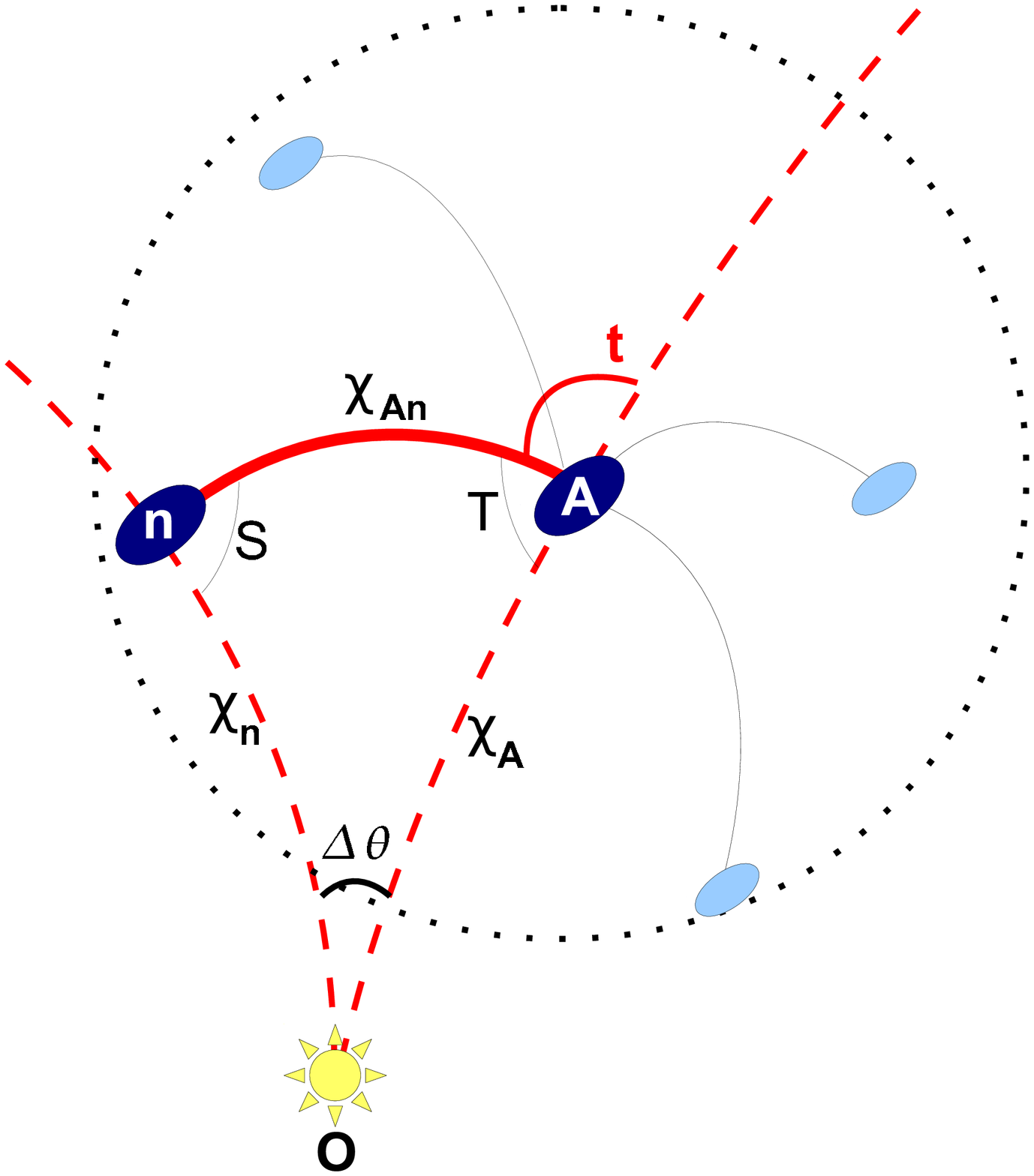}}
\centerline{\includegraphics[width=78mm,angle=0]{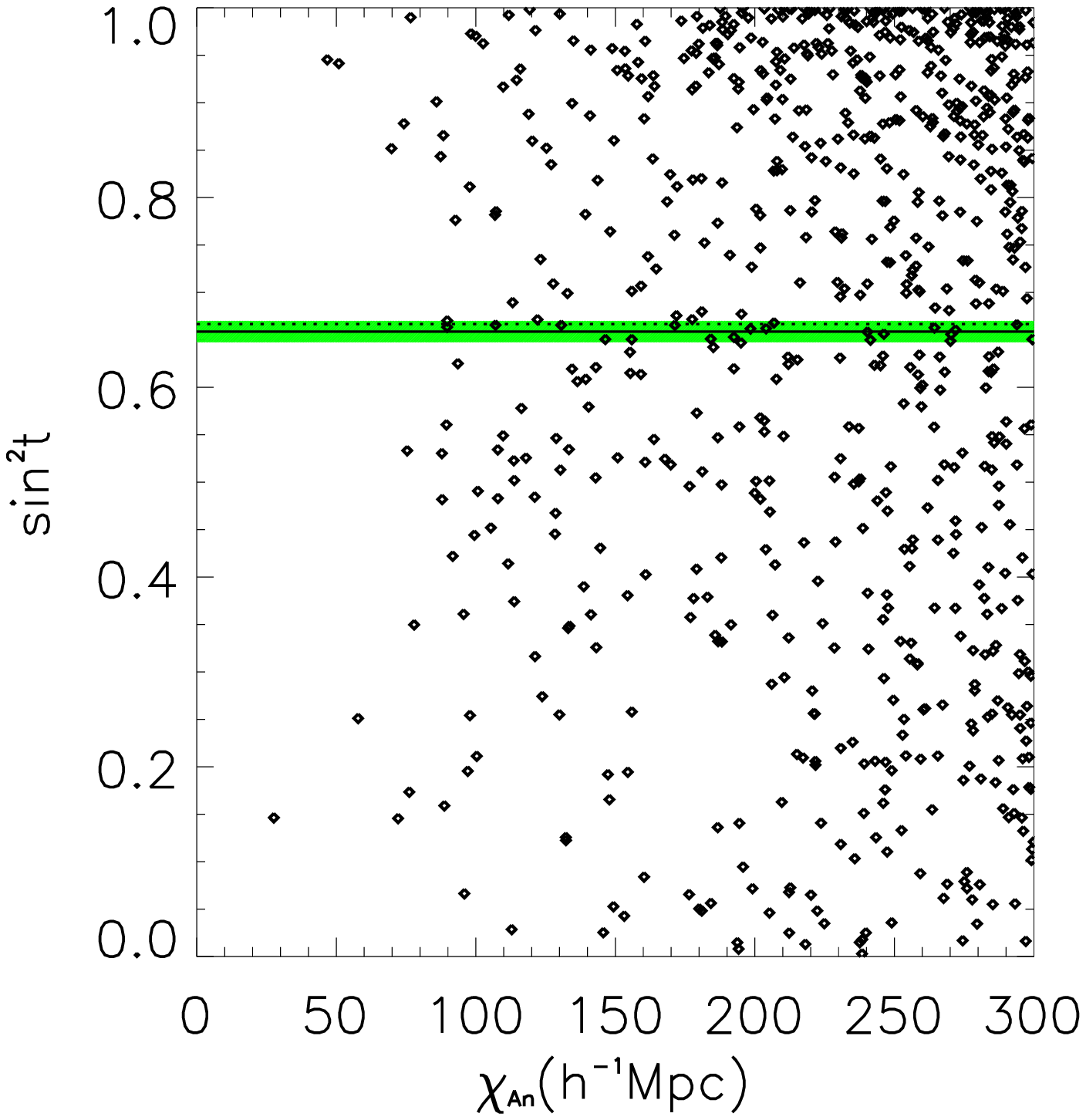}}
\caption{{\it Upper}
We determine the geodesic connections between a given target galaxy $A$ 
and all the surrounding galaxies that lie inside a sphere of radius $R$ centered on $A$. 
The target galaxy $A$ and its $n^{th}$ closest neighbor subtend an angle $\Delta \theta$ 
at the observer position $O$. The tilting angle $t$ measures the inclination of 
the geodesic separation $\chi_{An}$  between $A$ and $n$ with respect to the observer line-of-sight to $A$ (dashed line). If the CP holds on the scale $R$ we expect this {\it `spaghetti'} to be isotropically oriented about any given target. In other terms we expect the {\it los} angle $t$ to be isotropically distributed, i.e.\  its PDF is $\varphi(t) =(\sin t)/2$. 
{\it Lower:} the distribution of  $sin^2 t$ as a function of the 
geodesic separation $\chi_{An}$ in a sphere of radius  $R=300 h^{-1}$Mpc
randomly positioned in the LRG SDSS sample. 
The dotted line represents the theoretically expected average ($\mu=2/3$), while the solid line represents the IGI value, that is the average of the plotted points. 
The shaded area represents the $1\sigma$ uncertainty of the IGI value. 
}
\label{fig1}
\end{figure}

\[
ds^2=(cdt)^2-a^2(t)\big[d\chi^2+\Sigma_k^2(\chi) (d\theta^2+\sin^2\theta d\phi^2)  \big] 
\]

\noindent where, $c$ is the  light speed, $k$ is the scalar spatial curvature, $\chi$ is the radial geodesic comoving 
distance,  $a(t)$ is the cosmic expansion factor, 
and where, using the Kronecker symbol, 
$\Sigma_k(\chi)= \delta_{k,1} \sin \chi+\delta_{k,0}\chi+\delta_{k,-1}\sinh \chi$.
We  compute the length of the geodesic $\chi_{An}$ 
between the target galaxy $A$ and
its $n^{th}$ closest neighbour by exploiting the fact that  
the spatial part of the cosmic metric is  invariant under a 
{\it quasi-translation} transformation of its coordinates \cite{wey72}. 
We can thus translate the reference frame from the terrestrial 
observer $O$ to the target  $A$ and 
express the coordinates $\vec{x}_{n/A}$ of its $n^{th}$ neighbor as 
\begin{eqnarray}
\nonumber \displaystyle \vec{x}_{n/A} & = &\vec{x}_{n/O}-\vec{x}_{A/O} \Bigg\{ \left[ 1-k x_{n/O}^2\right]^{1/2}+ \\
&+&\left[ 1-(1-kx_{A/O}^2)^{1/2} \right] \frac{\vec{x}_{n/O} \cdot \vec{x}_{A/O}}{x_{A/O}^2} \Bigg\}\label{eq.quasitrslt}.
\end{eqnarray}

\noindent By orienting the axes in such a way to minimize the number of non-zero components 
(we choose $\vec{x}_{n/A}=\left(\Sigma_k(\chi_{An})\ ,\ 0,\ 0\right)$, 
$\vec{x}_{n/O}=\left(\Sigma_k(\chi_n)\sin \Delta \theta \ ,\ 0,\ \Sigma_k(\chi_n)\cos \Delta \theta \right)$ and $\vec{x}_{A/O}=\left(0\ ,\ 0,\ -\Sigma_k(\chi_A)\right)$),  
and by 
exploiting the identity 

\begin{equation} 
C_k^2(\chi)+k\Sigma_k^2(\chi)=1
\end{equation}
we obtain

\begin{eqnarray}
\label{eq.sigmaAn}
\Sigma_k^2(\chi_{An})&=&\Sigma_k^2(\chi_n)\sin^2\Delta\theta +\\
\nonumber &+&\Big[ \Sigma_k(\chi_n) C_k(\chi_A) \cos \Delta\theta-\Sigma_k(\chi_A)C_k(\chi_n)  \Big]^2.
\end{eqnarray}

\noindent 

Consider now the generalized law of sines \footnote{A straightforward way to obtain it
is by repeatedly applying equation (\ref{eq.sigmaAn})  to the 3 apexes of the 
triangle ($\hat{S},\hat{T}, \hat{\Delta\theta}$)  shown in Figure 1, and by isolating, 
after some algebra, identical terms in the resulting expressions.}

\begin{eqnarray}	
\frac{\sin \Delta \theta}{\Sigma_k(\chi_{An})}=\frac{\sin T}{\Sigma_k(\chi_{n})}=\frac{\sin S}{\Sigma_k(\chi_{A})}.
\label{lawsin}
\end{eqnarray}
\\*

\noindent Since $t=\pi-T$, it finally follows from eqs (\ref{eq.sigmaAn}) and (\ref{lawsin}) that 
\begin{eqnarray}
\sin^2t=\frac{1}{1+\left[C_k(\chi_{A})\ \cot \Delta \theta-\frac{\Sigma_k(\chi_{A})}{\Sigma_k(\chi_{n})}\frac {C_k(\chi_{n})}{\sin \Delta \theta}\right]^2}.
\label{sint}
\end {eqnarray}

If the CP holds, the {\it los} angle $t$  has a comoving space Probability 
Density Function (PDF) of a characteristic type ($\varphi(t)=(\sin t)/2$), 
namely, it is a random variable isotropically distributed with respect to any fundamental 
observer. Therefore, the expectation value $\mu=\langle \sin^2 t\rangle$ 
is cosmology independent and equal to $2/3$. 
We define the indicator of galaxy  isotropy (IGI) as the estimator $m_R$  constructed by averaging equation (\ref{sint}) over  
$n$ galaxies inside a sphere of comoving radius $R$ that is centered around any given observer in the universe.  
On scales $R$  where the CP applies, we expect the measure of  $m_R$ to converge to the predicted value  $\mu=2/3$ (see Figure 2).

The testing protocol is as follows: we assume that the CP holds and we implement
standard  statistical inference methods to try to falsify it and  reject its validity.
In detail,  we assume the existence of a  length-scale $R$ above which
the empirical IGI estimates ($m_R$) are statistically identical
to the theoretical prediction ($\mu$). We thus formulate a null hypothesis $h_0$ 
according to which the  two quantities are not different.
We quantify the goodness of the agreement by means of  
 $\chi^2$ statistics, and, following
standard convention,  we fix the rejection threshold of $h_0$, i.e. the risk of 
reaching the wrong conclusion,  at the $5\%$ level. 
This means that the hypothesis that the universe is isotropic above a scale 
$R$ cannot be rejected by data if the probability $P$ of obtaining a 
worst (larger) $\chi^2$ value is greater than 5\%.
On the contrary, an eventual failure in identifying the scale of isotropy would unambiguously point at the  incoherence of the  FLRW model.

Homogeneity and isotropy are  properties that characterize 
the large-scale distribution of {\it matter} on a 3D spatial hyper-surface
at a given instant of  time.  Since light propagates at a finite speed, the  most distant regions of the 3D volume
directly accessible to observations are also the furthest in time.
As a consequence, the number density of galaxies, 
an observable that is  modulated by local physical  processes with their own  
specific time-scales, is expected to vary  as a function of distance.
This is a known issue that  hampers most of the tests of the  CP \cite{hjm}.
In the following we show that, by focusing 
our attention on the angular distribution of galaxies, instead of their 
number density fluctuations, we can tackle the past light cone issue.  

If the CP holds true,  as we assume here, 
the  galaxy spatial number density $\rho_s(r)$ within  spherical shells of width 
$\Delta r$ 
centered on the terrestrial observer must be independent from the distance $r$. 
Note that {\it shell-homogeneity}, that is  $\rho_s=const$,  does not  
imply homogeneity,  i.e.\  invariance under  general spatial translations, while the 
opposite is true. More importantly,  the radial constancy of $\rho_s$ 
does not imply everywhere isotropy 
(isotropy about arbitrary comoving observers) that is the fundamental facet of the CP 
that we want to test. 
As a matter of fact, the distribution of galaxies that surrounds us 
can be characterized by a constant $\rho_s$ and yet be  anisotropic.
We therefore can legitimately remove past-light cone artifacts, 
by imposing that the comoving number 
density of galaxies be strictly constant within  concentric 
shells centered on us, without assuming homogeneity. 
In practice,  we analyze a volume limited catalog of galaxies, 
that is  a sample of objects brigther than a given  minimum absolute luminosity,  
and  we additionally remove, with a random rejection 
process, any residual radial gradient  in the distribution of galaxies. 
As we show in the Appendix, this technique  preserves the clustering properties 
of the galaxy distribution, and does not falsely impose homogeneity where 
there is none.

\begin{figure*}
\centerline{\includegraphics[width=175mm,angle=0]{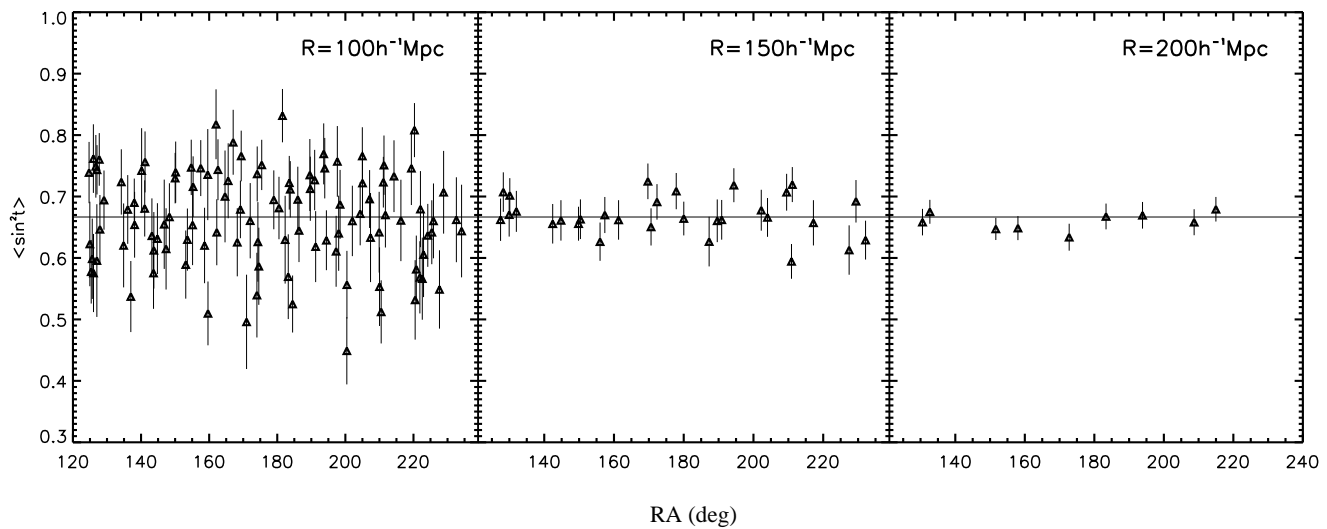}}
\caption{The IGI value  measured by different 
observers, labeled by their right ascension coordinate, is plotted. 
Each estimate is performed within non-overlapping spheres of comoving radius 
$R$ (shown in the inset) randomly thrown in the volume covered by the SDSS galaxy survey.
Errors are computed as  the standard deviation of the mean, and well trace the 
theoretically expected figure $\sigma=\sqrt{4/(45 n)}$. 
The average IGI value  ($m_R=\langle \sin^2 t\rangle$) is $0.678\pm0.005, 0.672\pm0.006$, and $0.660\pm0.007$ 
from left to right.  The solid line shows the expectation value 
predicted under the assumption that the  CP holds (i.e.\  $\mu =2/3$). 
On small scales data scatters widely,  while on 
scales where the CP is expected to hold, data fit the theoretical 
prediction. A goodness of fit statistical analysis yields $\chi^2/dof=(2.1, 1.12, 0.65)$ 
from the the smaller to the larger scale.} 
\label{pdf}
\end{figure*}

\section{Data}

We apply the method to the seventh release of the Sloan Digital Sky Survey \cite{dr7} which 
is comprised of $\sim 930,000$ galaxies over a field of view of 9380 deg$^2$.
Our analysis is limited to luminous red galaxies  (LRG, \cite{eise}) distributed in the North Galactic contiguous area defined by 
$120<RA<240$, $7<\delta<56$. 
A sample with a nearly constant density of galaxies  is obtained by 
volume limiting the SDSS dr7 catalog in the redshift range $0.22<z<0.5$.
This sample encompasses a comoving radial size   $\Delta r \sim 700h^{-1}$Mpc  (in what follows, we consider 
a cosmological model characterized by the reduced density of matter $\Omega_m=0.27$ and dark energy $\Omega_{\Lambda}=0.73$, and 
we assume that the value of the Hubble constant is $H_0=72 $ km s$^{-1}$Mpc$^{-1}$).
The upper redshift limit is fixed by the requirement of measuring the IGI with 
approximately the same average precision of nearly $1\%$ over all the interval  
$100<R <200h^{-1}$Mpc. 

The strict equality $\rho_s(r)=const$ is then imposed by 
 interpolating  the observed number density $\rho_s$ of objects 
in spherical shells centered on us (and  with thickness $\Delta r$ a hundred times smaller than the effective depth of the sample), 
and by randomly rejecting galaxies using a Monte Carlo process  with selection function 
$\phi(r)=\textrm{min}(\rho_s)/\rho_s(r)$. 
The final  LRG sample contains a total of $\sim 6500$ objects, has a mean  number 
density  $\rho=6.14 \cdot 10^{-6} h^3$Mpc$^{-3}$ and covers an 
effective field of view of $4860$ deg$^2$.

\section{Analysis of Data and Comparison to Theoretical Models}

For meaningful error interpretation, it is imperative to  acquire independent 
measurements of the IGI value, that is of the observable $m_R=\langle sin^2 t \rangle$ 
estimated using eq. (\ref{sint}). 
 Consequently, we do not apply our scheme to every  galaxy in the sample, i.e.\  we 
do not  carve  spheres of comoving  radius $R$ around each `extraterrestrial' observer  to determine whether they see the same degree of isotropy. 
Instead, we only select as observers, those target galaxies that are at  the center of non-overlapping spheres.    
As an example,  given the geometry of the  largest contiguous volume in the SDSS survey, we can place a maximum number $N=107, 30, 9, 4, 3$  of independent observers,    
exploring the isotropic distribution of galaxies on length-scales  $R=100, 150, 200, 250$ and $300 h^{-1}$Mpc respectively. 
Each of these observers are geodesically  connected, on average, to  $n=26, 87, 206, 401, 695$ galaxies respectively.

Before analyzing real data, we have first applied our method to  synthetic  samples simulating  spatially random (Poissonian)
galaxy distributions.  The point here is to detect the minimum radius $R_{iso}$ 
below which our technique is noise-limited and the scale of everywhere isotropy 
cannot be resolved.
Using Monte Carlo techniques, we have generated various uniform mock catalogues with galaxy number densities 
in the range $10^{-4} -10^{-6} h^3$Mpc$^{-3}$.
We have found that, as expected,  when the  scale $R$ is larger  than 
the mean inter-particle separation $\lambda=\rho^{-1/3}$,  the distribution of the $t$ angle statistically converges towards an isotropic PDF.
Quantitatively, as soon as  $R > 1.5 \lambda$, that is when on average $\sim 4\cdot(1.5)^3$ galaxies are geodesically connected to 
the observer,  the risk of reaching the wrong conclusion in rejecting 
the isotropy hypothesis $h_0$ becomes larger than $5\%$.
In particular, $R_{iso}$ of a spatially random  distribution of galaxies with the same density of the LRG sample investigated in this study 
can be unambiguously detected on scales larger than $\sim 85 h^{-1}$ Mpc.

We have then analyzed the LRG  galaxy sample extracted from the SDSS survey. 
The IGI value  estimated by distant observers on a scale $R=100,150$, 
and $200h^{-1}$Mpc  
is graphically shown in Figure 2.
It is interesting to note that, for any displayed scale $R$, 
the distribution of the average IGI values ($\langle m_R \rangle$) peaks
at $\mu=2/3$, while the variance of the distribution decreases as 
a function of $R$.  The stability of the central value of the distribution 
shows whether  isotropy is present on average, 
whereas the scatter shows whether isotropy is present for all observers.
In accordance with standard  theoretical expectations, 
as the $R$-scale increases, all the observers are equally likely to observe
isotropy, i.e.\  they lose their specificity and progressively become 
the `typical' observer of the universe. 
The upper panel of Figure 3 confirms that the galaxy pattern observed from
different positions in the universe approaches an isotropic distribution.   

The precise scale of transition to isotropy $R_{iso}$  is quantitatively determined as follows. First, by randomly rejecting galaxies from the main LRG sample, we have constructed  
1000 subsamples that satisfies  the requirement $\rho_s=const$. This bootstrapping process, allows us to estimate the central moments and  the dispersion of the $P$ statistics. 
We have then positioned the centers so that the maximum number of non-overlapping spheres of radius $R$ fit inside the survey volume. 
In particular, we require that the position of the extraterrestrial observers change randomly from sample to sample. 
For each length-scale $R$ probed, we have finally computed the risk  of erroneously 
rejecting the  null hypothesis as the median of $P$ over the 1000 realizations.  
The lower panel of Figure 3. shows that the  median risk is larger than $5\%$ for scales larger than 150$h^{-1}$ Mpc.
Despite the observed spread in the $P$-values for large $R$, essentially due to the low density 
of the LRG sample, a statistically significant sharp transition towards isotropy at a scale $R_{iso} \sim 150h^{-1}$ is unambiguously detected.  Notwithstanding, a 
larger  sample might also definitively exclude the hypothesis 
that the transition  happens at a scale as low as $120h^{-1}$Mpc. 
In fact, the evidence with which  such a low $R_{iso}$ is currently excluded
is  not yet conclusive.

As a matter of fact, 
the error with which the probability of such a
low transition scale is estimated from present data is not tightly constrained.

\begin{figure}
\centerline{\includegraphics[width=82mm,angle=0]{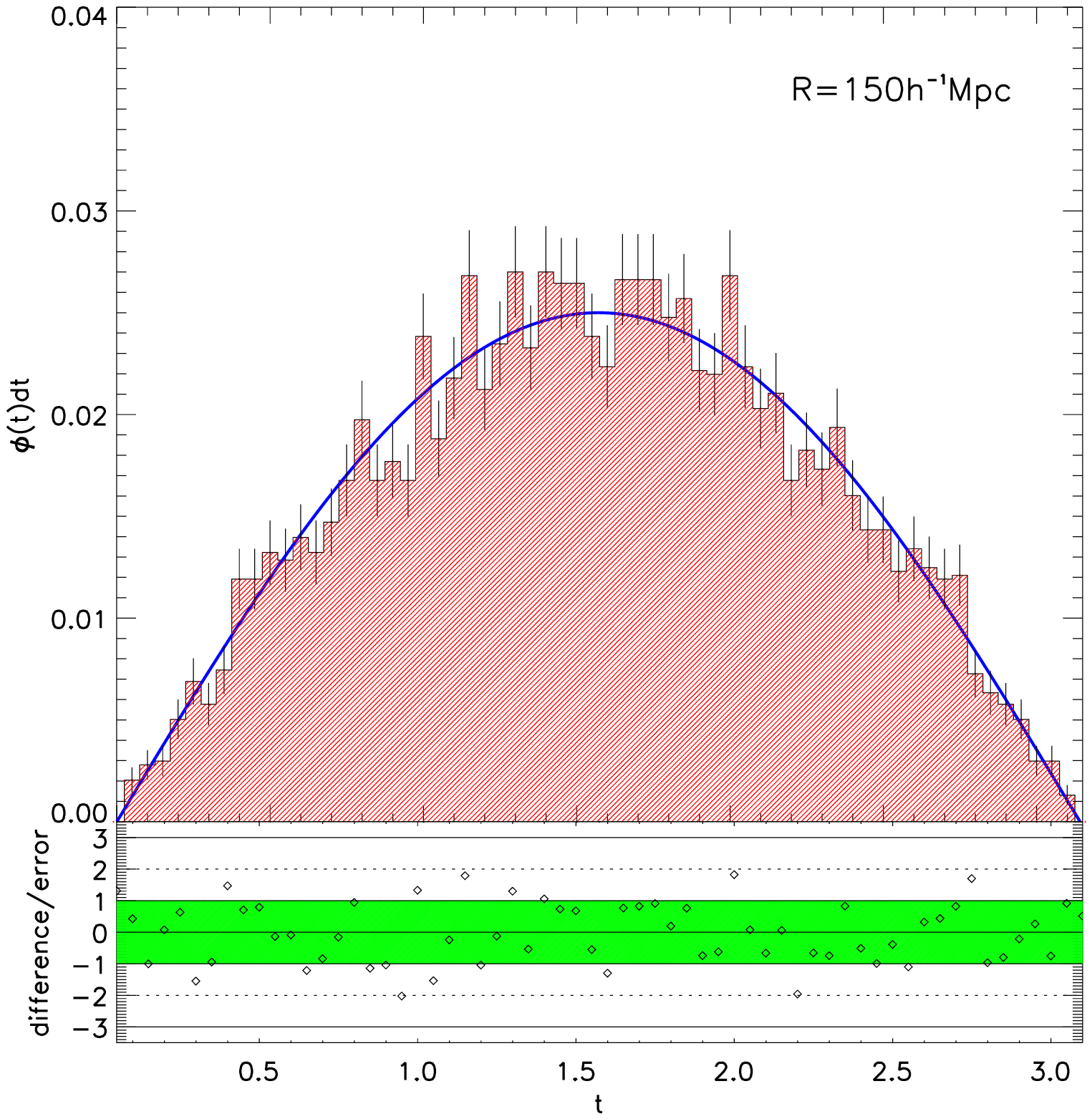}}
\centerline{\includegraphics[width=90mm,angle=0]{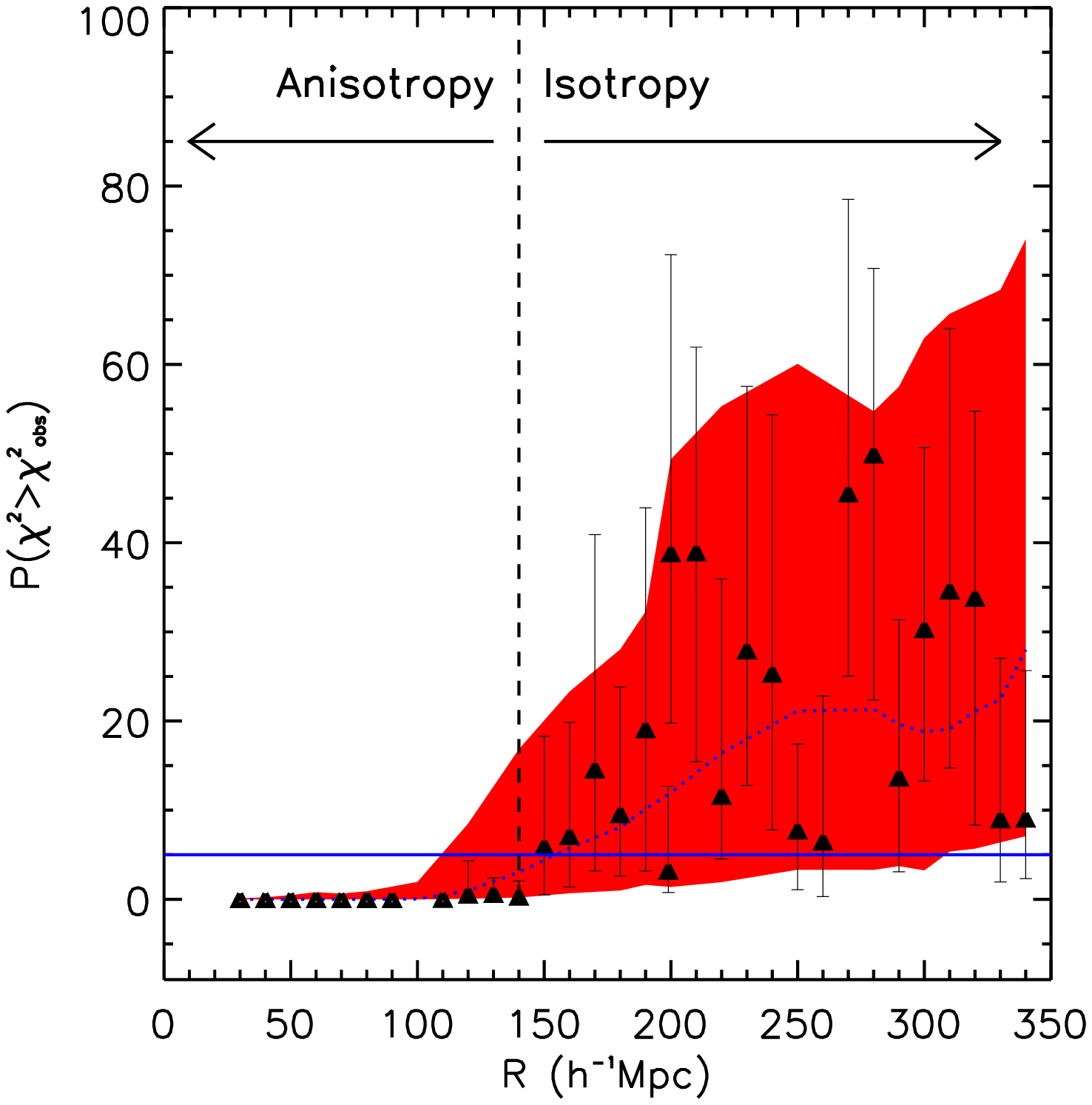}}
\caption{
{\it Upper:} the observed PDF of the {\it los} angle $t$ (histogram)
is compared to the isotropic prediction ($\varphi(t)=(\sin t)/2$)
on a scale $R=150h^{-1}$Mpc.
The ratio between  model deviations  and data errors is also plotted
(together with the lines indicating $1\sigma$ and $2\sigma$ deviations).
{\it Lower:} Dots show the levels of significance at which we fail to reject $h_0$
on a given scale $R$. We conventionally assume that the hypothesis of everywhere 
isotropy cannot be rejected when the risk of an erroneus decision is larger than 5\% 
(this threshold is represented by the solid blue line). The levels of significance 
are computed as the median of the probability $P$  
inferred from 1000  resamplings of the SDSS LRG sample that are  shell homogeneous. 
Error bars bracket the first and third quartile of the distribution of $P$ at any given 
scale $R$. The blue dotted line shows the average expectation for $P$ 
extracted from the analysis of 50 $\Lambda$CDM mock catalogs simulating the 
LRG SDSS sample. The red envelope shows the 
region bracketing 1$\sigma$ fluctuations around the average expectation extracted from 
simulations.}
\label{results.omol}
\end{figure}

Our conclusions are, within error bars,  independent of the density cut we 
artificially 
impose to guarantee shell homogeneity. We have verified that  samples that are everywhere isotropic on a scale $R_{iso}$ continue to be everywhere isotropic on that scale if 
the density threshold imposed by the requirements of shell homogeneity is enhanced. This conclusion follows from the analysis of subsamples obtained by   
volume-limiting the SDSS catalogue at a lower redshift $z_{max}$.  Interestingly, and in the opposite sense, if  the sample is isotropic on a scale $R_{iso}$ then 
it continues to be isotropic on that scale,  even when its density  is artificially lowered (by randomly rejecting galaxy members). 
The isotropic length proves robust  up until the investigated scale $R$ becomes smaller than 
$\sim 1.5$  times  the mean inter-particle separation. Below that threshold,  as  the analysis of random samples already suggested,  the predictive power  of our indicator breaks down.

We have compared our measurements to predictions of $N$-body 
simulations of the large-scale structure of the universe.
To this end, we have analyzed with the same technique  50 independent 
mock catalogs simulating the  distribution of LRG galaxies  in an 
SDSS-like survey. They were constructed by the LasDamas project \cite{damas}
using $\Lambda$ cold dark matter simulations (with characteristic parameters $\Omega_M = 0.25, \Omega_{\Lambda} = 0.75, h = 0.7, \sigma_8 = 0.8, n_s = 1$).

Figure 3 quantifies the confidence level with which the hypothesis $h_0$ cannot be 
rejected  on a scale $R$,  and compares it to what is expected in the mock catalogs.
Not only is a sharp transition towards isotropy at a scale $R \sim 150h^{-1}$Mpc  detected in real SDSS data,  it is  
observed  in synthetic galaxy catalogs too. This excellent agreement implies  that the scale of isotropy  $R \sim 150h^{-1}$  is a 
length that characterizes not only  luminous galaxies, i.e.\  the visible component of the universe, but also of the most massive dark matter halos.
The significance of this conclusion is best understood by considering that the everywhere isotropy  inferred from real data alone,  
does not give insight into the corresponding  arrangement of the underlying mass component.

\section{Conclusion}

An acritical acceptance of the Copernican principle might result in what Haynes \cite{Haynes}  called the ``Verrazzano bias".   
As in the case of this explorer  who, off the coast of the outer banks of North Carolina, mistakenly believed that he had discovered the Pacific Ocean, 
it is  dangerous  to  draw  definite  cosmological conclusions on the basis of limited data collected from a special spatial position. 

In this paper we have presented a new geometrical tool that allows us to assess whether or not,   from the view point of a distant 
galaxy,  the large-scale  structure of the universe appears almost identical to its aspect from earth.
Virtually all of the previous attempts to identify the coarse graining scale 
above which the visible distribution  of matter comply with the requirements 
of the CP have focused on the analysis of the so called {\it homogeneity scale}. 
In this paper we have addressed this same issue from a different angle. We propose
to identify this fundamental length with the scale of everywhere isotropy $R_{iso}$,
the scale above which the distribution of galaxies appears isotropic to every 
comoving observer, that we define as the smallest scale at which the probability 
of wrongly  rejecting the CP is smaller than $5\%$.

By analyzing state-of-the-art data, we have found 
that the galaxy distribution, as traced by luminous red galaxies,  appears isotropic 
to every comoving observer in the universe once the averaging scale 
is larger than $R_{iso} \sim 150 h^{-1}$ Mpc. 
This  figure is in excellent  agreement with predictions of the spatial 
clustering of galaxies in  $\Lambda$CDM simulations.

The advantage of the method is that it is insensitive to the
shape of the radial selection function of the redshift sample analyzed, i.e.\ 
to the effective number of objects that sample the underlying clustering of galaxies 
as a function of redshift. As a matter of fact, it is 
straightforward  to subtract  look-back time issues once the focus is shifted 
from counting objects (the standard methodology of the homogeneity tests)  to  
measuring angles (as implemented by our strategy). 

Since the matter distribution converges continuously towards homogeneity/isotropy, 
it is quite arbitrary to decide which criterium must be adopted to single out 
an exact scale of transition. In this work we adopt the point of view that 
the most natural way to test the CP is to 
assign a probability to the hypothesis that this model is wrong. 
The goal is to  frame the analysis of its  coherence within the domain of probability theory, 
as the intrinsically statistical nature of this cosmological statement  explicitly demands. 
This helps elucidating the meaning of such generic sentences as ``....the CP holds on a scales 
larger than $XXX$  Mpc" and will ease the quantitative comparison of the results obtained with different and independent methods.

Deeper redshift surveys of the universe (such as, for example,  BOSS, BigBOSS or EUCLID) are currently 
ongoing or expected to be soon completed. 
It would be interesting to understand if the scale of everywhere isotropy   
does scale as a function of cosmic time  as predicted by numerical simulations of the gravitational clustering in the universe.
This will confirm that the CP is not some temporary assertion about the present day appearance of the universe  but a fundamental 
property of matter distribution at all cosmic epochs. Even more importantly, it will help us  to shed light on the physics behind the 
large-scale uniformity of the universe by answering the question:  
where does this scale come from?

 \vskip 1.truecm

 \noindent {\bf Acknowldgements}.
We would like to thank A. Blanchard, P.S. Corasaniti,  L. Moscardini, E. Scannapieco,  
T. Schucker,  P. Taxil and S. Troubetzkoy  for useful discussions. 
We thank the anonymous referee for  in-depth comments and  valuable suggestions 
that helped us to improve the presentation of the manuscript.
C.M. is grateful for support from specific project funding of the {\it Institut  Universitaire de France}.


\section{Appendix A}

The fair sample model of the universe (Layzer 1956), assumes that 
the galaxy distribution  is a discrete stochastic process resulting from 
the Poissonian sampling of an underlying continuous matter density 
field $\Lambda({\bf x})$. Accordingly, a galaxy sample  
that traces  an underlying continuous field of PDF $Q(\Lambda)$
can be modeled as  a discrete stochastic process in which the 
probability of counting $N$ galaxies within a given arbitrary cell 
is  

\begin{equation}
P_N=\int P(N|\Lambda)Q(\Lambda)d\Lambda, 
\label{pn}
\end{equation}

\noindent where  the conditional sampling probability is 

\begin{equation}
P(N|\Lambda)=\frac{\Lambda^{-N}}{N!} e^{-\Lambda}.
\label{pois}
\end{equation}

In the ideal universe described by the FLRW model, the continuous matter fluid 
$\Lambda({\bf x})$ satisfies exactly  the CP and its PDF is  the 
Dirac delta $Q(\Lambda)=\delta^{D}(\Lambda-\Lambda_0)$.
Equations (\ref{pn}) and (\ref{pois}) therefore imply that  $P_N$ 
follows a Poissonian statistic, that is  the spatial  
distribution of galaxies is random.

We now demonstrate that a spatially random  sample of size $K$ cannot be generated by 
randomly sampling, with probability $p$,  a parent population 
of size $N$ whose spatial distribution is  clustered, that is inhomogeneus.
Let the sampling process be described by

\begin{equation}
K=\Phi({\bf x}_1)+\Phi({\bf x}_2)+.....+\Phi({\bf x}_N),
\label{sam}
\end{equation}

\noindent where $\Phi$ is a random variable (taking on the values 0 or 1) 
distributed according to the Bernoulli probability law 
(that is $p^{\Phi}(1-p)^{1-\Phi}$,  where $p$ is the probability 
that $\Phi=1$ in a single trial), and where $N$ is a non-negative integer-valued 
random variable  distributed according to the Poisson distribution of average 
counts $\bar{N}$.

The probability generating function (PGF) of $P_N$ is 
\begin{equation}
\mathcal{G}(z)\equiv \sum_{i=0}^{\infty} P_i z^{i},
\label{defG}
\end{equation}
 
\noindent and its expression in the case of the Bernoulli and Poisson processes is 

\begin{equation}
\mathcal{G}_{\Phi}(z)=(1-p)+pz
\end{equation}
\noindent and
\begin{equation}
\mathcal{G}_{N}(z)=e^{(z-1) \bar{N}},
\label{gpois}
\end{equation} 

\noindent respectively.

It can be shown \cite{kest} that the PGF 
of the sum of independent, identically distributed random variables $\Phi$, 
i.e.\ the PGF of the random variable $K$ in  equation (\ref{sam}) is 

\begin{equation}
\mathcal{G}_{K}(z)=\mathcal{G}_{N}\big(\mathcal{G}_{\Phi}(z)\big).
\label{theo}
\end{equation}

Let's assume that $\mathcal{G}_K$ is the PGF  of a Poissonian distribution 
with average parameter $\bar{n}$. By taking the derivative of eq. (\ref{theo}) 
we obtain 

\begin{equation}
\bar{n} e^{\big[\frac{(\mathcal{G}_{\Phi}+p -1)}{p} -1\big]\bar{n}}=\frac{d\mathcal{G}_N}{d \mathcal{G}_{\Phi}} p
\end{equation}

\noindent which, upon integration gives

\begin{equation}
\mathcal{G}_{N}(u)=e^{(u-1)\frac{\bar{n}}{p}}+C.
\end{equation}

\noindent 

The value of the arbitrary constant  $C$  can be set to $0$
using the additional condition  $\mathcal{G}_{N}(1)=1$ (see eq. (\ref{defG})).
Therefore,  a spatially random distribution cannot be the result of the random 
sampling of a non-Poissonian distribution.  
Reciprocally, one can show that a random sampling 
of a Poisson parent distribution, results in a subsample of elements 
with Poisson distribution. 

We now show that also the $N$-point moments of the continuous mass
distribution $\Lambda ({\bf x})$ are not modified by a random sampling process.
Given the counts $N_i\equiv N({\bf x}_i)$ in a cell at position ${\bf x}_i$,
the $N$ point PGF is immediately obtained by generalizing  the expression given
in eq. (\ref{defG})

\begin{equation}
G(z_1, z_2, ....,z_n)=\sum_{N_1} \sum_{N_2} ...\sum_{N_n} P_{N_1,N_2,...N_n} z_1^{N_1} z_2^{N_2}....z_n^{N_n},
\end{equation}

\noindent where 
\begin{equation}
P_{N_1,N_2,...,N_n}=\int P(N_1|\Lambda_1)....P(N_n|\Lambda_n))Q(\Lambda_1)...Q(\Lambda_n) d\Lambda_1.....d\Lambda_n,
\end{equation}

\noindent and where  $\Lambda_i \equiv \Lambda({\bf x}_{i})$.
If the sampling is random,  the $N$-point PGF  is given by

 \begin{equation}
G(z_1, z_2, ...., z_n)= \langle e^{\Lambda_1(z_1-1)} e^{\Lambda_2(z_2-1)}......e^{\Lambda_n(z_n-1)}\rangle.
\end{equation}

The moment generating function associated to the discrete, $N$-point, counts $N_1, N_2, ....,N_n$  follows immediately by substituting $z_i \rightarrow e^{z_i}$ in the argument of the PGF \cite{szasza}

\begin{equation}
M(z_1, z_2, ....,  z_n) = \langle e^{\Lambda_1 (e^{z_1}-1)} e^{\Lambda_2(e^{z_2}-1)}, ...., e^{\Lambda_n(e^{z_n}-1)}\rangle.
\label{chf}
\end{equation}

The $N$-point moment of the galaxy distribution,  is calculated as the functional 
derivative  of the characteristic function \cite{bmar}

\begin{equation}
\langle N({\bf x}_1)N({\bf x}_2).....N({\bf x}_n) \rangle \equiv \frac{\delta^{n}M}{\delta z_1 \delta z_2....\delta z_n}\Big |_{z_1=z_2=......=z_n=0}.
\end{equation}

 By substituting  eq. (\ref{chf}) into the previous one, we
see that  the $N$-point galaxy moment is equivalent to the corresponding statistics 
computed for the underlying continuous matter field, i.e.\ 
\begin{equation}
\langle N({\bf x}_1)N({\bf x}_2).....N({\bf x}_n) \rangle =
\langle  \Lambda({\bf x}_1)\Lambda({\bf x}_2).....\Lambda ({\bf x}_n)\rangle.
\end{equation}

As a result, the whole hierarchy of N-point galaxy correlation functions 
computed from a random (discrete) sample trace with fidelity the $N$-point correlation 
function of matter.  Again, an inhomogeneous spatial distribution cannot be turned 
into a homogeneous one by a random sampling process.
\end{document}